\documentclass[aps,prl,superscriptaddress,twocolumn]{revtex4-1}

\usepackage{bm}
\usepackage{slashed}
\usepackage{url}

\begin{document}


\title{Dark energy and mass generation}


\author{Marco Frasca}\email{marcofrasca@mclink.it}
\affiliation{via Erasmo Gattamelata, 3, 00176 Roma, Italy}


\date{\today}

\begin{abstract}
We consider a set of solutions for a massless quartic scalar field, recently devised, that satisfy a massive dispersion relation. We show that such solutions have the property to give the correct behavior for the equation of state of the dark energy. It is seen that conformal invariance is restored and the mass gap goes to zero on a time scale determined by the Hubble constant and the strength of the self-interaction of the scalar field. When conformal invariance is restored, the equation of state for the dark energy can apply.
\end{abstract}


\maketitle

\section{Introduction}

One of the more striking results in cosmology, recently acquired, is that the universe expansion is accelerating. This effect can be described by adding a cosmological term to the Einstein equations \cite{Peebles:2002gy,ame,lap}. One of the ways one generally uses to understand this effect is by a scalar field pervading all the universe termed ``quintessence''. There is not a general way to give the exact equation for the self-interaction term of this field and so, there are also formulations that just avoid it.

After the recent discovery of the Higgs particle at the LHC \cite{Aad:2012tfa,Chatrchyan:2012ufa}, it would be really an interesting matter if the same mechanism would be at work for yielding masses and accelerating the universe. But the current Higgs model is unsatisfactory for such an application.

Recently, we devised a new set of solutions for a massless quartic scalar field \cite{Frasca:2009bc} that describes a field with a massive dispersion relation. Such solutions are proven to be the starting point for a quantum field theory in the strong coupling limit \cite{Frasca:2013tma}. This theory, having a mass gap and so a non trivial vacuum expectation value at the leading order, can be used to describe mass generation in a different way with respect to the Higgs mechanism.

In this letter we aim to study the behavior of these solutions at the classical level as a quintessence field. We will show that the behavior is that expected for the dark energy. Indeed, the equation of state shows a negative pressure and the density of energy has the proper order of magnitude one expects for the cosmological term.

\section{Exact solutions}

The equation for the quintessence field $Q$ has the general form
\begin{equation}
   \ddot Q+3H\dot Q+V'(Q)=0
\end{equation}
where $H$ is the Hubble constant and $V(Q)$ the potential with the prime meaning a derivative with respect to $Q$ and the dot is the derivative with respect to time. For a general scalar field on a flat space-time it is possible that a mass gap develops. Indeed, let us consider the case of a quartic scalar field as
\begin{equation}
   \partial^2\phi+\lambda\phi^3=0
\end{equation}
being $\lambda$ the coupling. An exact set of solutions is yielded by \cite{Frasca:2009bc,Frasca:2013tma}
\begin{equation}
\label{eq:exeq}
   \phi(x)=\mu\left(\frac{2}{\lambda}\right)\mathrm{sn}(p\cdot x+\theta,-1)
\end{equation} 
with $\mu$ and $\theta$ two integration constants and sn a Jacobi elliptic function, provided that the following dispersion relation holds
\begin{equation}
\label{eq:dr}
    p^2=\mu^2\sqrt{\frac{\lambda}{2}}
\end{equation}
that appears to be that of a massive field. The corresponding quantum field theory, with such a field pervading all the space-time, maintains the mass gap and adds a 
Kaluza-Klein spectrum on the free particles. This theory has a trivial infrared fixed point \cite{Frasca:2013tma}. We are going to show that such solutions have the property to behave as a quintessence field with the equation of state proper to the cosmological term in the Einstein equations.

Our technique will be based on the fact that the ratio $H/m_0$, being $m_0=\mu(\lambda/2)^\frac{1}{4}$ is greatly lesser than unity and so, it can be used as a development parameter. Then, we firstly exploit the behavior of the theory at the leading order.

\section{Partition function \label{app1}}

It is possible to evaluate the partition function for the scalar field at the infrared fixed point. We can use the set of classical solutions described in the preceding section. So, let us suppose to have such classical solutions $\phi_c(x)$ and study the generating functional around them taking $\phi(x)=\phi_c(x)+\delta\phi(x)$, being $\delta\phi(x)$ the fluctuations around the classical value. We will give these solutions explicitly below taking into account the periodicity condition $\phi({\bm x},0)=\phi({\bm x},\beta)$. Now, by a direct substitution one gets
\begin{equation}
   Z=Ne^{-S_0[\phi_c]}\int[d\delta\phi]e^{-S_1[\delta\phi]}
\end{equation}
being
\begin{eqnarray}
   S_0&=&\int_0^\beta d\tau\int d^3x
   \left[\frac{1}{2}
   \left(\frac{\partial\phi_c}
   {\partial\tau}\right)^2
   +\frac{1}{2}(\nabla\phi_c)^2+\frac{\lambda}{4}\phi_c^4\right] \nonumber \\
   S_1&=&\int_0^\beta d\tau\int d^3x
   \left[\frac{1}{2}\left(\frac{\partial\delta\phi}{\partial\tau}\right)^2
   +\right. \nonumber \\
   &&\left.\frac{1}{2}(\nabla\delta\phi)^2+\frac{3}{2}\lambda\phi_c^2(\delta\phi)^2\right].
\end{eqnarray}
We have turned back to a Gaussian integral that we are able to solve as we know the propagator of the linear equation for the fluctuations. Anyhow, we just evaluate the leading order contribution. We assume a finite volume $V=L_xL_yL_z$ and the Matsubara frequencies changed to  $\omega_{k_\tau}=4k_\tau K(-1)/\beta$, being $K(-1)$ the complete elliptic integral of the first kind, while for the other momenta in a box we have $p_l=4k_\tau K(-1)/L_l$ with $l=x,y,z$. The exact classical solutions $\phi_c(x)$ we are working with can be written down as in eq.(\ref{eq:exeq}) with $\theta=0$, with the dispersion relation (\ref{eq:dr}) provided to hold. Then, the leading order contribution to the partition function is, ignoring an irrelevant constant,
\begin{equation}
   Z={\rm Tr}\left(\exp\left[-\beta V\frac{\mu^2}{2}\sqrt{\frac{2}{\lambda}}
   \left(p_0^2+\frac{1}{3}{\bm p}^2\right)\right]\right)
\end{equation}
where the remaining trace is on all the momenta. in principle, we have to sum on all $n_k$ but we take the continuum limit and write down
\begin{equation}
   Z_0=\beta V\int\frac{d^4p}
   {(2\pi)^4}\exp\left[-\beta V\frac{\mu^2}{2}\sqrt{\frac{2}{\lambda}}
   \left(p_0^2+\frac{1}{3}{\bm p}^2\right)\right]
\end{equation}
and finally
\begin{equation}
   Z_0=\frac{3^\frac{3}{2}}{8\pi^2}\frac{\lambda}{\mu^4}\frac{T}{V}
\end{equation}
from which we can derive the equation of state as
\begin{equation}
   P=T\frac{\partial}{\partial V}\ln Z_0= -\frac{T}{V}
\end{equation}
that is negative representing the pressure of the vacuum at temperature $T$ at the infrared fixed point. Correspondingly, the energy is given by
\begin{equation}
   U=-\frac{\partial}{\partial\beta}\ln Z_0= T
\end{equation}
and is positive. From these results is clear that we are coping with an ensemble of free particles as one should expect from the trivial infrared fixed point of this theory. But what is more interesting here is the behavior of the pressure for these solutions that appears to be negative. Then, we will use such a behavior of the scalar field at low energies to analyze it as a model for dark energy. This is so because we can rewrite the equation of state as
\begin{equation}
   p=-\frac{\langle E\rangle}{V}=w\rho
\end{equation}
with $\rho$ the energy density and $w=-1$ as expected for dark energy. For a temperature of about 2.7~K and the estimated volume of the universe, this expression gives a proper order of magnitude for the cosmological constant.

\section{Dark energy equation of state}

Once such exact solutions are given, we can evaluate the equation of state for dark energy. For this aim we just consider the rest frame with
\begin{equation}
\label{eq:exeq0}
   \phi(t,0)=\mu\left(\frac{2}{\lambda}\right)\mathrm{sn}(m_0 t+\theta,-1)
\end{equation}
We just note that the pressure is
\begin{equation}
   p=\frac{1}{2}\dot\phi^2-\frac{\lambda}{4}\phi^4
\end{equation}
and the density is
\begin{equation}
   \rho=\frac{1}{2}\dot\phi^2+\frac{\lambda}{4}\phi^4.
\end{equation}
So,
\begin{equation}
  w=\frac{p}{\rho}=\frac{\frac{1}{2}\dot\phi^2-\frac{\lambda}{4}\phi^4}{\frac{1}{2}\dot\phi^2+\frac{\lambda}{4}\phi^4}.
\end{equation}
This yields
\begin{equation}
  w=\frac{p}{\rho}=1-2\mathrm{sn}^4(m_0 t+\theta,-1).
\end{equation}
This result shows that, if the mass gap goes to zero and conformal invariance is restored, we get $w=1-2\mathrm{sn}^4(\theta,-1)$. So, depending on the phase $\theta$, we get back the equation of state for the dark energy.

\section{Hubble constant}

So far we have worked with the approximation $H=0$. This approximation holds provided the mass gap is much larger than the Hubble constant, that is $\mu(\lambda/2)^\frac{1}{4}\gg H$. We can consider a perturbation series having this ratio as a development parameter. Our aim here is to compute the next to leading order correction. The equation to solve is
\begin{equation}
    \ddot\phi+\lambda\phi^3=-3H\dot\phi
\end{equation}
being $H$ the Hubble constant. We rescale time using the mass gap as $\tau=m_0 t$ with $m_0=\mu(\lambda/2)^\frac{1}{4}$. Similarly we put $\hat\phi=\phi/m_0$. We arrive at the equation
\begin{equation}
    \ddot{\hat\phi}+\lambda\hat\phi^3=-3\epsilon\dot{\hat\phi}
\end{equation}
being $\epsilon=H/m_0$ our development parameter for the perturbation series. Taking $\hat\phi=\hat\phi_0+\epsilon\hat\phi_1+\epsilon^2\hat\phi_2+\ldots$ one has
\begin{eqnarray}
    &&\ddot{\hat\phi_0}+\lambda\hat\phi_0^3=0 \nonumber \\
		&&\ddot{\hat\phi_1}+3\lambda\hat\phi_0^2\hat\phi_1=-3\dot{\hat\phi_0} \nonumber \\
		&&\ddot{\hat\phi_2}+3\lambda\hat\phi_0^2\hat\phi_2=-3\hat\phi_0\hat\phi_1^2-3\dot{\hat\phi_1} \nonumber \\
		&&\vdots.
\end{eqnarray}
So, the next to leading order solution takes the form, after undoing rescaling,
\begin{equation}
    \phi_1=-H m_0 t\mathrm{sn}(m_0 t,-1)-\frac{H}{2}m_0^2t^2\mathrm{cn}(m_0 t,-1)\mathrm{dn}(m_0 t,-1).
\end{equation}
This correction has secular contribution running as polynomials to infinity at increasing time. This secularities can be removed using a renormalization group techniques as devised in \cite{Kunihiro:1995zt}. To do this, we rewrite the approximate solution at a generic initial time $t_0$ and reintroduce the phase at $\theta(t_0)$. We will have
\begin{eqnarray}
    &&\phi(t,t_0)=\mu(t_0)\left(\frac{2}{\lambda}\right)^\frac{1}{4}\mathrm{sn}(m_0 t+\theta(t_0),-1) \nonumber \\
		&&-\epsilon m^2_0(t-t_0)\mathrm{sn}(m_0 t+\theta(t_0),-1) \nonumber \\
		&&-\frac{\epsilon}{2}m_0^3(t-t_0)^2\mathrm{cn}(m_0 t+\theta(t_0),-1)\mathrm{dn}(m_0 t+\theta(t_0),-1) \nonumber \\
		&&+O(\epsilon^2)
\end{eqnarray}
and calculate the envelope deriving with respect to $t_0$. One has
\begin{eqnarray}
   \left.\frac{d\phi}{dt_0}\right|_{t_0=t}=0
\end{eqnarray}
and so ($\epsilon=H/m_0$)
\begin{eqnarray}
   &&\frac{d\bar\mu}{dt}=-H\sqrt{\frac{\lambda}{2}}\bar\mu+O(\epsilon^2) \nonumber \\
	 &&\frac{d\theta}{dt}=O(\epsilon^2).
\end{eqnarray}
This yields a time-dependent mass shift as $\bar\mu(t)=\mu e^{-H\sqrt{\frac{\lambda}{2}}t}$ at the next to leading order. This is to decay with time and so, the mass gap will go to zero as time increases. Physically this means that the conformal invariance of the scalar field will be restored on a time scale determined by the Hubble constant and the strength of the self-interaction. In turn, this implies that, on the given time-scale, one recovers the equation of state of dark energy.

\section{Conclusions\label{conc}}

We have shown that mass generation can be strictly connected to a quintessence field producing dark energy of the universe. This effect is due to the mass gap arising from the self-interaction of the scalar field. The mass gap is seen to go to zero on a time scale determined by the product $H\sqrt{\frac{\lambda}{2}}$ and so, the Hubble constant can be responsible for the restoration of the conformal invariance and the dark energy we observe today.

\end{document}